\documentclass[fleqn,10pt]{wlscirep}
\usepackage[utf8]{inputenc}
\usepackage[T1]{fontenc}
\usepackage{layouts}
\usepackage{setspace}
\usepackage{longtable}
\usepackage[version=4]{mhchem}       
\usepackage{multirow}
\usepackage{array}
\usepackage{hyperref}
\usepackage[version=4]{mhchem}

\usepackage{xr}
\usepackage{cleveref}

\definecolor{light-gray}{rgb}{0.9,0.9,0.9}

\newcolumntype{P}[1]{>{\centering\arraybackslash}p{#1}}

\title{Wired Beneath, Scented Above: A Dual-Parameter Approach to Soil-Plant Interaction via Electrical Resistance and VOC Analysis}

\author[1, *]{Mridul Kumar}
\author[2]{Krishnananda Soami Daya}

\affil[1]{Department of Materials Engineering, Ben-Gurion University of the Negev, Be'er Sheva, Israel}
\affil[2]{Microwave Physics Lab, Department of Physics and Computer Science, Dayalbagh Educational Institute, Agra, India}

\affil[*]{Corresponding author email: mridul@post.bgu.ac.il}

\keywords{Plant Communication, MQ-3 Sensors, Spearman Correlation}

\begin{abstract}
Plants, being sessile organisms, have evolved a variety of defence mechanisms to protect themselves from invaders such as pathogens, insects, and herbivores. One key strategy is the release of volatile organic compounds (VOCs) into the air, which serve as warning signals to nearby plants, prompting them to activate their own defence mechanisms. Although plant communication through VOCs is well-documented, the interaction between VOC emissions and soil-released allelochemicals remains less understood. In this study, we investigated the relationship between the release of allelochemicals and VOCs by chickpea (gram) plants, focusing on their role in plant signalling. We grew 25 chickpea plants in individual beakers with soil as the growth medium. The release of allelochemicals was monitored indirectly by measuring the electrical resistance of the soil, while VOC emissions were analysed using MQ-3 gas sensors. Continuous monitoring provided insights into soil chemistry changes potentially influenced by allelochemical release, alongside the VOC profile captured by the sensors. Spearman correlation analysis was applied to evaluate the relationship between allelochemical release and VOC emissions. Understanding this interplay could contribute to the development of sensor networks for early detection of plant stress, offering a potential strategy for reducing crop losses and improving agricultural resilience.
\end{abstract}

\begin{document}

\flushbottom
\maketitle
%
%
\thispagestyle{empty}


\section*{Introduction}

Plant communication refers to the ways in which plants exchange information and interact with each other, other organisms, and their environment. While plants lack the nervous system and sensory organs found in animals, they have evolved a range of complex mechanisms to communicate and respond to various stimuli. Plants release VOCs, which are airborne chemicals that can serve as signals to nearby plants. These compounds can convey information about herbivore attacks, pathogen infections, and environmental stress. When one plant is attacked, it can release VOCs that warn neighbouring plants to activate defence mechanism \cite{dudarevaBiosynthesisFunctionMetabolic2013}. 

Literature from the early 1970s demonstrates that plants have a self-defence mechanism and use chemical defences against infestation from pathogens, insects or weeds \cite{schoonhoven1972secondary, rhoades1976toward, kumarDecodingPhysiologicalResponse2023}. In self-defence mode, plants absorb more nutrients from the soil and produce phenolic compounds these are called allelochemicals (such as tannins, lignins and phenol derivatives) which also act as toxins to insects and animals. Studies show that these chemicals can change the digestibility of plant leaves \cite{berensBalancingTradeoffsBiotic2019,wangCriticalRolePotassium2013}. It is also reported that different insects respond differently to these allelochemicals produced by plants \cite{feeny1976plant}. 

According to Rhoades (1985), physical damage to trees can affect the quality of the leaves of undamaged tree leaves \cite{rhoades1985offensive}, these claims were also earlier observed by Baldwin \textit{et al}., where poplar (\textit{Populus euroamericana}) ramets and sugar maple (\textit{Acer saccharum Marsh}) seedlings exhibited a high concentration of phenol derivatives in leaf extracts within 75 hours of having the leaves of neighbouring plants physically cut \cite{baldwinRapidChangesTree1983}. Furthermore, many species of small herbivores like insects, rabbits, and leaf eating birds exhibit wide variations in population numbers through time. These variations are of great interest to pest managers. Conventionally, large variation in population was attributed to competition for food, weather-induced mortality and natality of species and differing rate of predation, disease and parasitism (Cold Spring Harbor Symposia on Quantitative Biology, 1957). But, even in the abundance of food, no predation, parasitism or disease, population collapses have been reported \cite{rhoades1985offensive}. This variation in population remained a mystery until the behaviour of plants towards herbivores was explored.

In traditional models, plants were seen as passive participants in their relation with herbivores. There are strong evidences that plants are far from being passive \cite{rhoades1985offensive}. Plants initiating chemical responses Under stress (biotic or abiotic stress) is one of these. It is also observed that these chemical responses keep the plants defended against herbivores or herbivorous insects. Based on this model, it was proposed that the variation of population in herbivores is due to allelochemicals that they were consuming along with the leaves of plants. This proved that plants not only feel their surroundings but can also respond and react to changes, just like a living species. 

Karban \textit{et al.}, reviewed 48 well-replicated studies on plant behaviour, and they observed that plant resistance towards external damage (herbivory or insects) increased for individuals with damaged neighbours \cite{karban2014volatile}. This observation concludes a signalling mechanism in plants. This signalling makes undamaged individuals to change their traits and become more resistant towards incoming danger. One of the collected studies was from 1983, in which it was reported that Sitka willow (\textit{Salix sitchensis}) when grown with damaged neighbours shows a higher level of resistance to herbivores \cite{rhoades1983responses}. This signalling between plants happen because of sharing of volatile organic compounds (VOCs) and generally referred to as "plant communication". In most of the cases the VOC was found to be ethylene gas \cite{farmer2001surface}. 

Volatile cue sharing or communication is sometimes referred to as "eavesdropping" by the plants on other plants because in actual communication information is exchanged between two or more parties. However, here only one plant is emitting the signal in the form of VOCs and other plants are listening to this signal \cite{bradbury2000economic}. In earlier studies, plant communication has been seen only in ideal lab conditions, where plants had been kept in tight jars and containers, therefore, had also been criticised a lot \cite{dickePlantsTalkAre2003}. However, published literature of last decade voids all the criticism by carefully studying the plant communication in its well constructed and replicable studies, experimentalists have emphasised on plants using VOCs for sending the warning signs to nearby plants, when stressed \cite{karban2013kin,karban2014volatile}. 

Some studies suggest that root might also be the means for communication in plants \cite{mahall1991root,mahall1992root}, while in another study root to shoot communication has also been seen \cite{blackman1985root}. An experiment on desert shrubs \textit{Ambrosia dumosa} and \textit{Larrea tridentata} showed results about the behaviour of roots \cite{mahall1991root}. Their results concluded that inter-root detection requires a contact or diffusion of chemicals over short distances. The reduction shown in root elongation in \textit{Ambrosia} after interplant root contact might be related to some kind of self-nonself detection mechanism \cite{burnet1971self}. This self-nonself identification mechanism helps plants in preventing antagonistic behaviour towards its own root branches \cite{karban2009self}.

Despite these advances, the mechanisms underlying these complex signalling processes such as the interplay between the VOCs and allelochemicals is still not fully understood. For instance, it has been reported that abiotic stress in the root causes shootward transport of stress signals to induce stress responses on the whole-plant level \cite{KO2017R973,blackman1985root} which leads to the release of VOCs in the air for warning the nearby plants about the danger \cite{falik2011rumor}. While VOCs help manage interactions above ground, alerting neighbouring plants and attracting beneficial insects, allelochemicals influence below-ground interactions by modulating soil chemistry and microbial communities. Together, these mechanisms form an integrated communication and defence network that enhances the plant's ability to adapt, compete, and survive in its environment. Targetting both of these stress response mechanisms can help in early detection of the plant stress \cite{kumarDecodingPhysiologicalResponse2023}. 

In this study, we assessed the release of allelochemicals by chickpea (gram) plants in soil using electrical resistance measurements, while simultaneously analysing the VOC profile with MQ-3 gas sensors. The objective was to explore the relationship between allelochemicals released in the soil and VOC emissions in the air. We grew 25 chickpea plants in individual beakers with soil as the growth medium, continuously monitoring the electrical resistance of the soil. The surrounding VOC profile was recorded using MQ-3 sensors. Spearman correlation analysis revealed a strong correlation ($ \ge 0.75$) among most plant pairs, suggesting communication that led to similar electrical resistance patterns. Additionally, a moderate correlation ($\ge 0.40$) was observed between the electrical resistance and VOC profile across most plant - MQ-3 sensor pairs. Notably, as the distance between plant pairs increased, the correlation in their electrical resistance profiles decreased, indicating that the strength of inter-plant communication diminishes with distance.

\section*{Results}

\subsection*{Experiment 1 : Comparison of Soil Behaviour with Agarose}

\subsubsection*{Comparison of the Electrical Resistance of Soil with Agarose}
\begin{figure}[h]
\centering
\includegraphics[width=15cm]{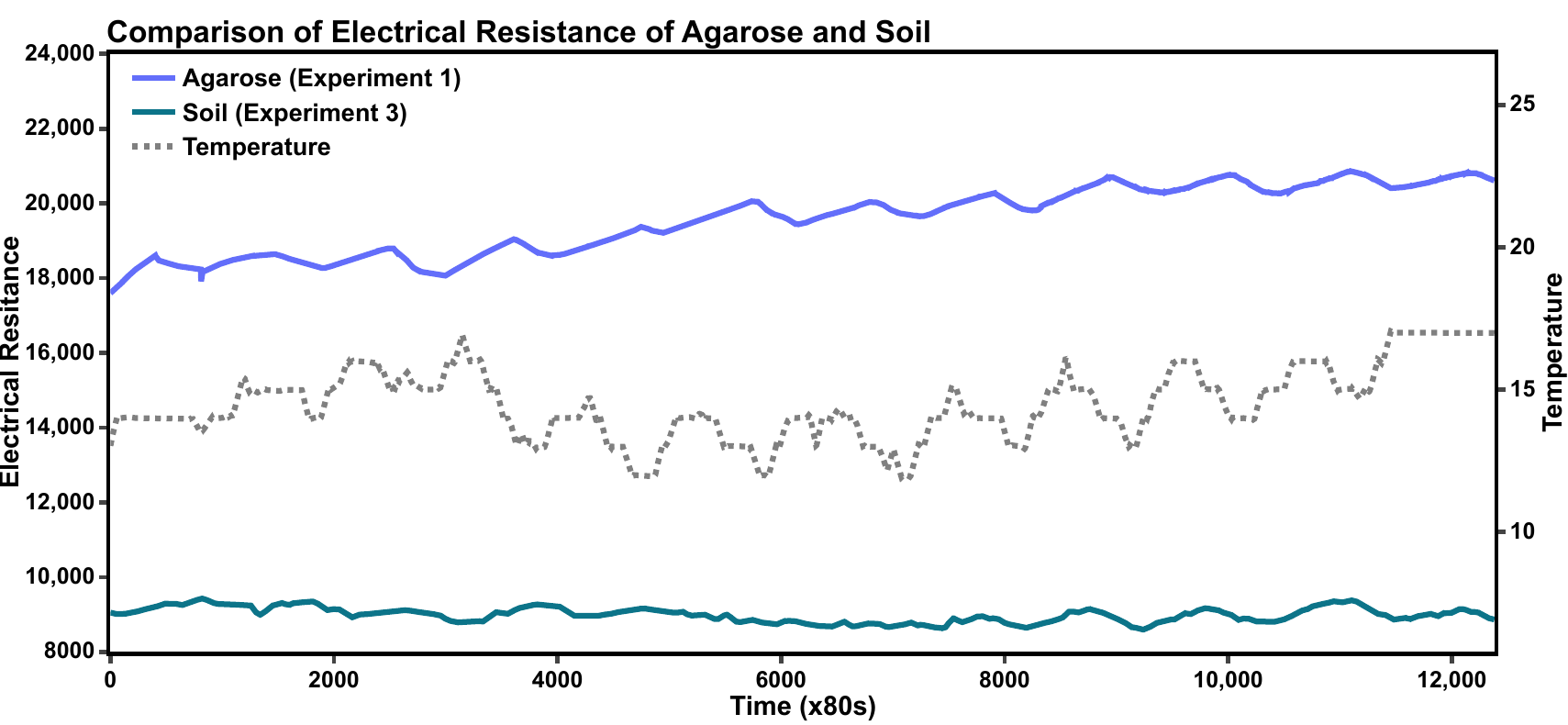}
\caption{Comparison between the electrical resistance of agarose growth media to electrical resistance of soil in bottle 8 during experiment 1. From the graphs, it can be seen that in both of the cases the electrical resistance characteristics is quite similar.}
\label{fig:er_comparison}
\end{figure}

After cleaning the electrical resistance of the soil from noise (see Supplementary Section \ref{supp-sec:noise_cleaning} and Supplementary Figure \ref{supp-fig:noise_cleaning}), it was compared with the electrical resistance property of the agarose growth media. On graphing the electrical resistance of the soil in 7$ ^{th} $ beaker (it only had soil, so pure electrical resistance of soil will be seen) and agarose growth media (see Figure \ref{fig:er_comparison}), it was observed that electrical resistance of agarose is almost twice that of soil which can be attributed to the fact that the soil contains more nutrients than agarose. It can also be observed that the periodic variation of the electrical resistance of soil is similar to that of agarose and both of the curves show a positive correlation coefficient of 0.68. The periodic variation in electrical resistance of soil shows that it also has a temperature compensation factor (TCF) and electrical resistance is inversely proportional to the temperature similar to agarose.

\subsubsection*{Temperature Compensation Factor of the Soil}

Even if soil is semi-solid in nature, it contains minerals which in the presence of water remain in their ionic forms \cite{alvaRELATIONSHIPIONICSTRENGTH1991,bradyNaturePropertiesSoils2002,brevikSoilElectricalConductivity2016}. This leads to a TCF for the soil \cite{kimIonChannelbasedFlexible2018}. If $ \alpha $ is the TCF of soil $ R $ and $ T $ are the arbitrary electrical resistance and temperature respectively, $ R_0 $ and $ T_0 $ are the reference electrical resistance and temperature then TCF is given by, \cite{kumarDecodingPhysiologicalResponse2023},
\begin{equation}
	\alpha = \frac{R_0 - R}{R(T - T_0)}
	\label{eq:temp_comp}
\end{equation}

From the Supplementary Figure \ref{supp-fig:temp_res} the values of $ R $, $ T $, $ T_0 $, and $ R_0 $ have been extracted and listed in Table \ref{tab:dataset}. From these values, TCF has been calculated using equation \ref{eq:temp_comp}. The calculated value of the TCF for soil comes out to be 0.0271 $ ^\circ C ^ {-1} $. It is to be noted that TCF depends on several factors such as soil type, concentration of water in soil etc. \cite{brevikSoilElectricalConductivity2016}.

\begin{center}
	\setlength{\tabcolsep}{7pt} 
	\renewcommand{\arraystretch}{1.2} 
	\begin{longtable}{|c|c|c|c|c|}
		\caption[]{Values of \boldmath{$ R $}, $ T $, $ T_0 $, and $ R_0 $ both while increasing temperature and decreasing temperature.}
		\label{tab:dataset}  \\
		\hline \multicolumn{1}{|c|}{\boldmath{$ R_0 (\Omega)$}} & \multicolumn{1}{c|}{$T_0 (^\circ C)$} & \multicolumn{1}{c|}{\boldmath{$R (\Omega)$}} &  \multicolumn{1}{c|}{\textbf{T ($^\circ$C)}} & \multicolumn{1}{c|}{\boldmath{$\alpha = \frac{R_0 - R}{R(T - T_0)}$ ($ ^\circ C^{-1} $)}} \\ \hline 
		\endfirsthead
		
		\hline \multicolumn{1}{|c|}{\boldmath{$ R_0 (\Omega)$}} & \multicolumn{1}{c|}{\boldmath{$T_0 (^\circ C)$}} & \multicolumn{1}{c|}{\boldmath{$R (\Omega)$}} &  \multicolumn{1}{c|}{\textbf{T ($^\circ$C)}} & \multicolumn{1}{c|}{\boldmath{$\alpha = \frac{R_0 - R}{R(T - T_0)}$ ($ ^\circ C^{-1} $)}} \\ \hline
		\endhead
		\hline 
		\endfoot
		\multicolumn{5}{|c|}{\textbf{Decreasing Temperature}} \\ \hline
		8997 & 12.44 & 9438 & 10.95 & 0.0314 \\ \hline
		8800 & 14    & 9275 & 12.9  & 0.0466 \\ \hline
		8642 & 14.44 & 9138 & 12    & 0.0222 \\ \hline
		8604 & 13.96 & 9173 & 11.94 & 0.0307 \\ \hline
		8748 & 14.31 & 9384 & 12.33 & 0.0342 \\ \hline
		
		\multicolumn{5}{|c|}{\textbf{Increasing Temperature}} \\ \hline
		9119 & 12.2  & 8769 & 14.43 & 0.0179 \\ \hline
		9252 & 12.02 & 8956 & 14    & 0.0167 \\ \hline
		9143 & 11.96 & 8591 & 14.1  & 0.03   \\ \hline
		9173 & 11.94 & 8748 & 14.31 & 0.0205 \\ \hline
		9384 & 12.33 & 8886 & 15.04 & 0.0207 \\ \hline
		
		\multicolumn{4}{|c|}{\textbf{Mean}} & \textbf{0.0271} \\ \hline
\end{longtable}
\end{center}

\subsection*{Experiment 2 : Eavesdropping on Plant Communication}
Experiment 1 shows the similarity between the electrical resistance properties of agarose growth media and soil, however, soil has a higher TCF because of the presence of higher amount of nutrients. Therefore, the experimental setup used in experiment 1 can be translated to the real world conditions using soil as a growth media \cite{kumarDecodingPhysiologicalResponse2023}. Therefore, we move on to conduct experiment 2 which had two iterations. 

\subsubsection*{Iteration 1 of Experiment 2}

\begin{figure}[h]
	\centering
	\includegraphics[width=12cm]{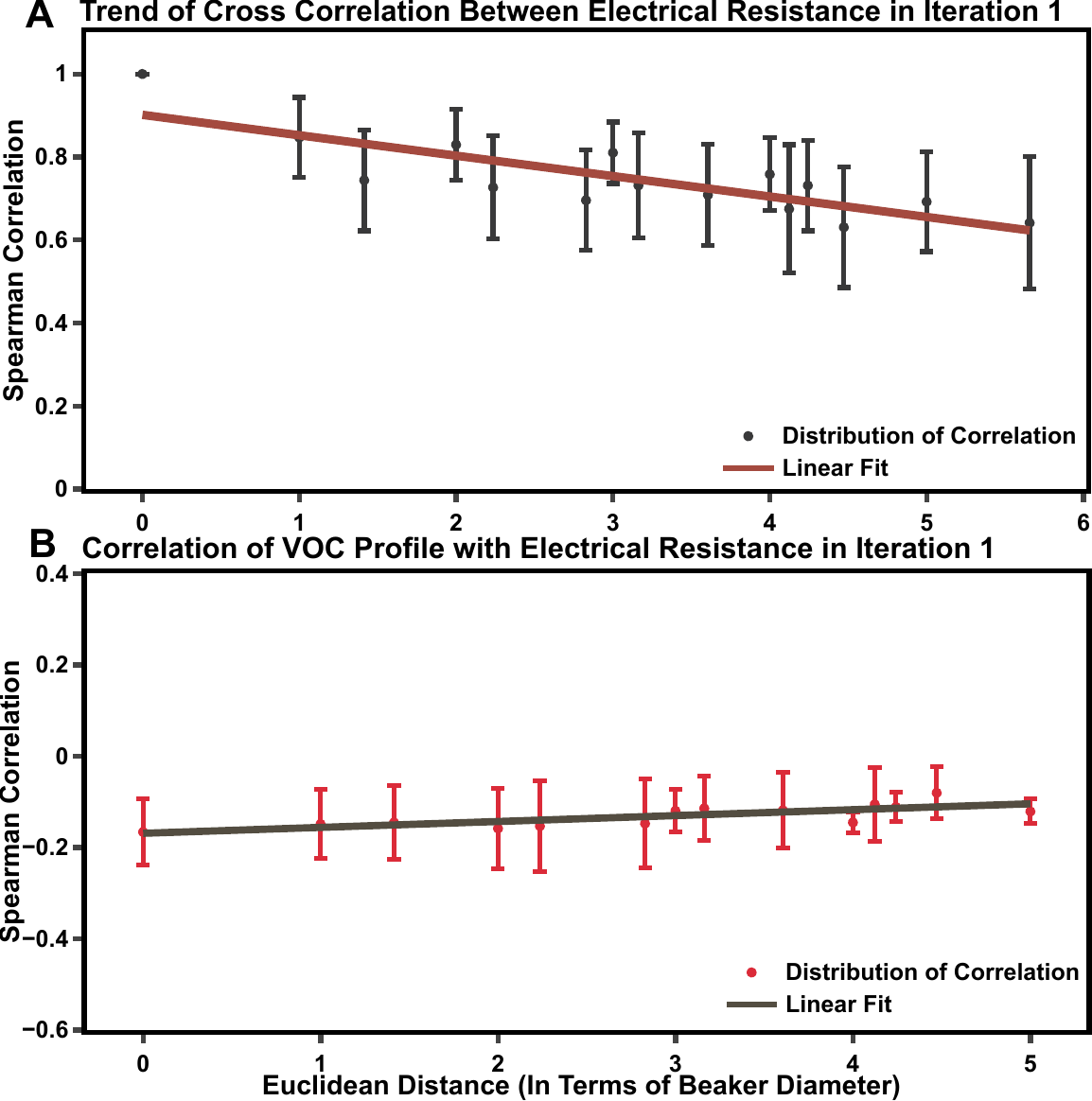}
	\caption{A. The variation of the cross correlation between the electrical resistance values of the plants in the beakers with respect to the distance. It can be seen that the correlation overall decreases indicating that the plant communication is distance dependent. B. This figure shows the variation of the correlation between the electrical resistance and the VOC profile of the plants. The correlation also tends to 0 meaning as the distance increases, plant communication ceases to exist.}
	\label{fig:er_stats_i1}
\end{figure}

\begin{figure}
	\centering
	\includegraphics[width=12cm]{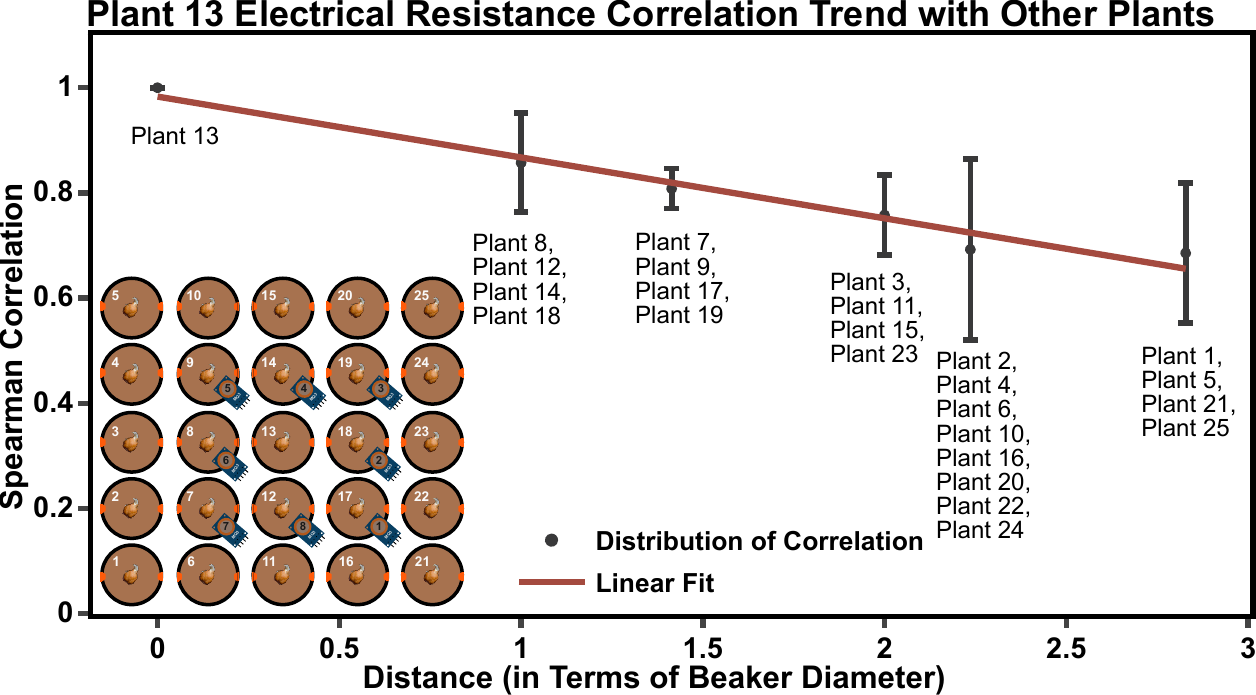}
	\caption{Correlation trend of electrical resistance of plant 13 with the other plants and its variation with the distance.}
	\label{fig:plant13_it1}
\end{figure}

The primary objective of this iteration is to understand correlation between electrical resistance readings of soil in different beakers and the VOC profile of the plants. Therefore, Spearman Correlation Coefficient (SCC) was calculated between electrical resistance readings and VOC profile of the plants and also the electrical resistance characteristics of one plant to the other (see Supplementary Figures \ref{supp-fig:corr_er_i1} and \ref{supp-fig:corr_er_i2} and Supplementary Figures \ref{supp-fig:corr_voc_i1} and \ref{supp-fig:corr_voc_i2}). The variation of SCC was also studied with the distance (see Figure \ref{fig:er_stats_i1}A and \ref{fig:er_stats_i1}B). It can be observed that the electrical resistance values of the plants are highly correlated with each other, with an SCC greater than 0.75 (see Supplementary Figure \ref{supp-fig:corr_er_i1}). It can also be observed that the position of the beakers has a significant impact on SCC between the electrical resistance characteristic of two plants. SCC overall decreases as the distance between the two plants increases (see Figure \ref{fig:er_stats_i1}A). It is to be noted that since the beakers are arranged in 5 $\times$ 5 arrangement, a Euclidean distance is the correct measure of distance. For plant 1, SCC shows a downward trend to plant 5 and then increases with plant 6 because it is closer to it (see Supplementary Figure \ref{supp-fig:corr_er_i1}). Plants 22-25 are away from plant 1 and show a low SCC with it. Similarly, for plant 13, which was grown at the centre of the arrangement, shows decreasing SCC with other plants as the distance increases (see Figure \ref{fig:plant13_it1} and Supplementary Figure \ref{supp-fig:it1_plant13}).

If we look into the SCC between plant's electrical resistance characteristics with the VOC data from MQ-3 sensors, a negative SCC is observed for most plant – MQ-3 sensor pairs with the exceptions of $4^{th}$ MQ-3 sensor which shows a positive correlation for 15/25 plants, $ 2^{nd} $ MQ-3 sensor which shows positive SCC with plant 3 and 4, and $ 8^{th} $ MQ-3 sensor which also shows positive correlation with plants 3 and 4. When plants are stressed they release VOCs in the air and allelochemicals in the soil which are acidic in nature with high electrical conductivity \cite{batishPhenolicAllelochemicalsReleased2007,dasAllelopathicPotentialitiesLeachates2012}. Therefore, the electrical resistance of the soil would decrease. This indicates that there would be a negative correlation between electrical resistance and VOC profile generated from the MQ-3 sensors, and our data shows the same.

\subsubsection*{Iteration 2 of Experiment 2}

In iteration 2 of the experiment 2, we observed that few plants (1, 2, 3, 4, 6, 8, 9, 14, 15, 16, and 20) did not show a significant growth. Therefore, these plants were removed from further analysis. The complete correlation data can be found in the supplementary file.

\begin{figure}
	\centering
	\includegraphics[width=12cm]{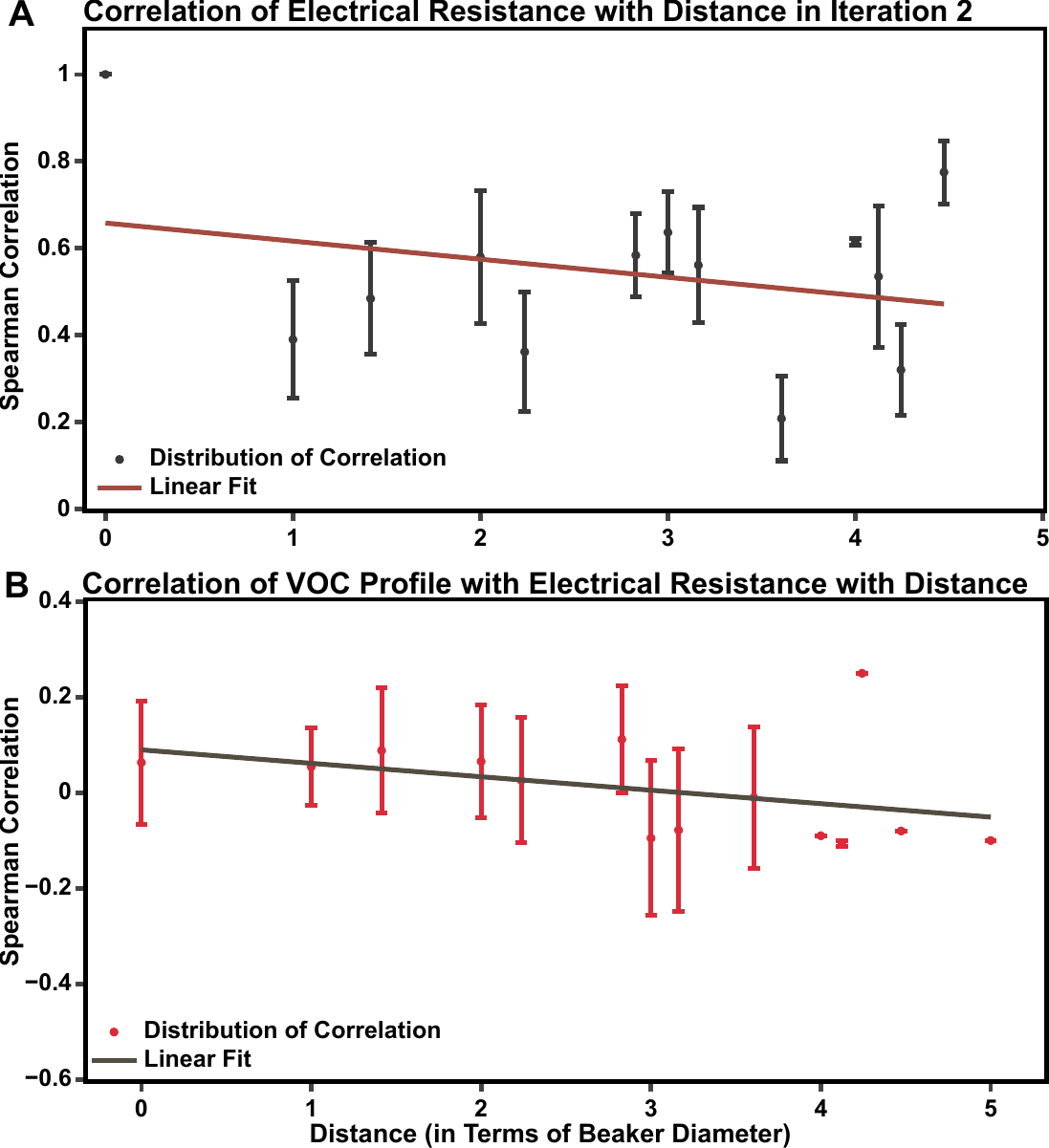}
	\caption{A. The trend of the cross correlation of the plant electrical resistance with the distance between the plants. B. The trend correlation between the VOC profile of the plants with the electrical resistance with distance between the plant and the sensor}
	\label{fig:it2_stats}
\end{figure}

\begin{figure}
	\centering
	\includegraphics[width=15cm]{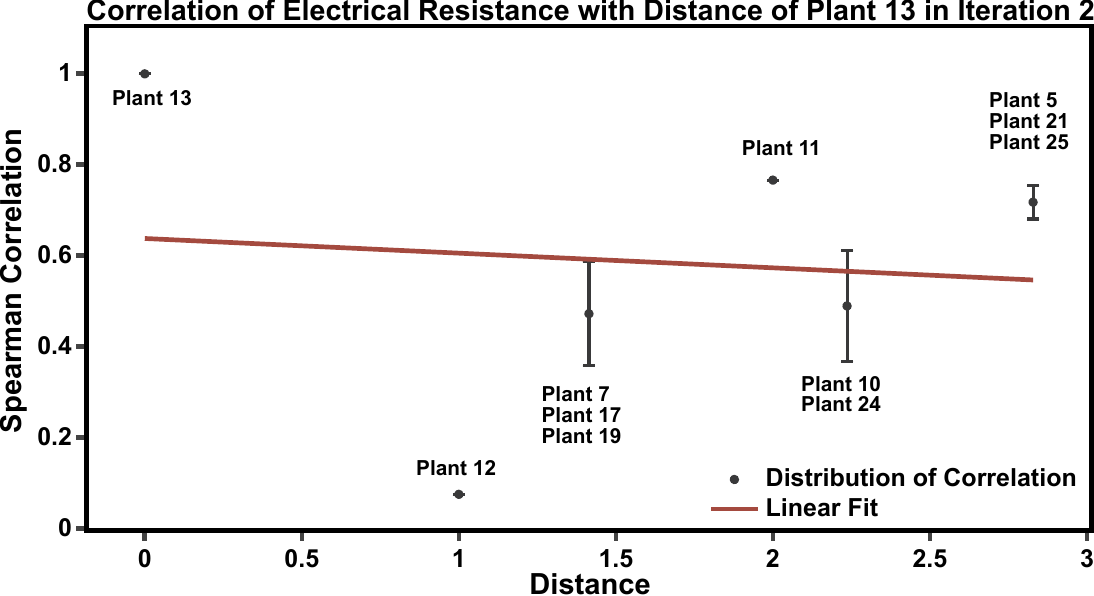}
	\caption{The variation of correlation of plant 13's electrical resistance with the other plants with respect to the distance.}
	\label{fig:plant13_it2}
\end{figure}

Similar to the iteration 1, we observe that there is an overall negative slope of the variation of cross correlation of plant's electrical resistance with the distance (see Figure \ref{fig:it2_stats}A). However, in iteration 2 the absolute value of the correlation coefficient has decreased which can be attributed to the fact that few plants did not show significant growth which has led to the overall decrement in the VOC concentration in the environment \cite{karban2014volatile}. Plant 13 also shows an overall negative trend of the electrical resistance correlation with the other plants with increased distance, however, plant 12 shows is an exception (see Figure \ref{fig:plant13_it2}).

The correlation of the electrical resistance with the VOC profile shows a negative trend and moves towards 0 as the distance between the sensor and the plant increases (see Figure \ref{fig:it2_stats}B). However, contrary to the previous iteration, plants near the sensor show a positive correlation between the electrical resistance and the VOC profile measured by the sensor. As the distance between the sensor and the plants increases the values turn negative with an exception of plant 21 with sensor 6.

\section*{Discussion}

Experiment 1 compares the electrical properties of the agarose growth media and soil. For this purpose, electrical resistance of the soil has been compared with that of the agarose with a moderately strong correlation coefficient of 0.68. However, soil shows a higher value of the temperature compensation factor (0.0241/$^\circ C$) which can be associated with the higher amount of the nutrients in it as compared to agarose \cite{bradyNaturePropertiesSoils2002}.

Iteration 1 of Experiment 2 analyzed 300 unique plant-plant pairs. Among these, only 0.67\% (2 pairs) exhibited a negligible correlation coefficient ($|corr| \le 0.25$). A weak correlation ($0.25 \le |corr| \le 0.50$) was observed in 12.6\% of the pairs, while 29\% showed a moderately strong correlation ($0.50 \le |corr| \le 0.75$). Notably, 57.6\% of the pairs demonstrated a strong correlation ($|corr| > 0.75$) (see Supplementary Figure \ref{supp-fig:corr_er_i1}). These results suggest a form of communication between the plants, likely influencing their electrical resistance characteristics to develop similarly because of the initiation of their self defence mechanism \cite{farmer1990interplant,baldwinRapidChangesTree1983}.

When examining the correlation between the electrical resistance of plants and their VOC (Volatile Organic Compound) profiles, 200 unique MQ-3 sensor-plant pairs were analyzed. Of these, 25\% had a correlation coefficient between -0.1 and 0.1, indicating minimal correlation. Additionally, 52.5\% showed a weak correlation ($0.1 \le |corr| \le 0.2$), 13\% had a moderate correlation ($0.2 \le |corr| \le 0.3$), 5.5\% displayed a stronger correlation ($0.3 \le |corr| \le 0.4$), and 4\% reached the highest examined range of ($0.4 \le |corr| \le 0.5$) (see Supplementary Figure \ref{supp-fig:corr_voc_i1}).

Similarly, when we examine the correlation coefficients between the electrical resistance characteristics for iteration 2 of the experiment 2, we observe that for 18.6\% of plant - plant pairs show negligible correlation coefficient ($ |corr| \le 0.25 $), 24\% pairs show weak correlation coefficient ($ 0.25 \le |corr| \le 0.50 $), 39\% pairs show moderate correlation ($ 0.50 \le |corr| \le 0.75 $), and 18.3\% pairs show the strong correlation ($ |corr| > 0.75 $) (see Supplementary Figure \ref{supp-fig:corr_er_i2}). In this iteration the number of plant pairs showing negligible correlation has increased which can be attributed to the fact that some of the plants did not show significant growth which leads to the decrement in the amount of VOCs released in the environment around the plants. Therefore, the communication between the plants is not as significant as in iteration 1.

When we examine the correlation between the electrical resistance and the VOC profile of the plants we observe that out of 200 unique plant - MQ3 sensor pairs 39.5\% show negligible correlation ($-0.1 \le |corr| \le 0.1$), 27\% pairs showed weak correlation ($0.1 \le |corr| \le 0.2$), 16\% pairs showed moderate correlation coefficient ($0.2 \le |corr| \le 0.3$), 12.5\% pairs showed strong correlation ($0.3 \le |corr| \le 0.4$), and only 4\% could show a highest correlation of $|corr| > 0.4 $ (see Supplementary Figure \ref{supp-fig:corr_voc_i2}). The previously stated fact that some of the plants did not grow and showed minimal correlation between each other can also be verified from this data as significant number of plants ($\sim 40\%$) do not show correlation with VOC profile measured with MQ-3 sensors which is higher than the iteration 1 in which only 25\% plant - MQ3 sensor pairs did not show significant correlation.

\section*{Methods}

This study has been divided in two experiments namely 1 and 2. Experiment 1 has been designed to translate the previously used experimental setup \cite{kumarDecodingPhysiologicalResponse2023} to the real world conditions i.e. by using the soil as a growth media. The primary purpose of conducting this experiment was to see the similarity between the electrical resistance properties of the soil to the agarose growth media. While on the other hand experiment 2 was designed to study the relationship between the volatile organic compounds (VOCs) profile and the electrical resistance of the growth media i.e. soil. 

Soil is a heterogeneous mixture of organic matter, minerals, gases, liquids, and a myriad of organisms that support plant life \cite{bradyNaturePropertiesSoils2002}. The relative proportions of these components determine the type of soil, which can range from nutrient-rich loam to nutrient-poor sand. Soil can vary greatly in composition and structure, depending on factors such as climate, vegetation, and geology. These factors also affect the concentration of nutrients in the soil. Depending upon the concentration of nutrients in the soil, electrical resistance would be decided.

\subsection*{Design and Working of Experimental Setup}

\begin{figure}
	\centering
	\includegraphics[width = 15cm]{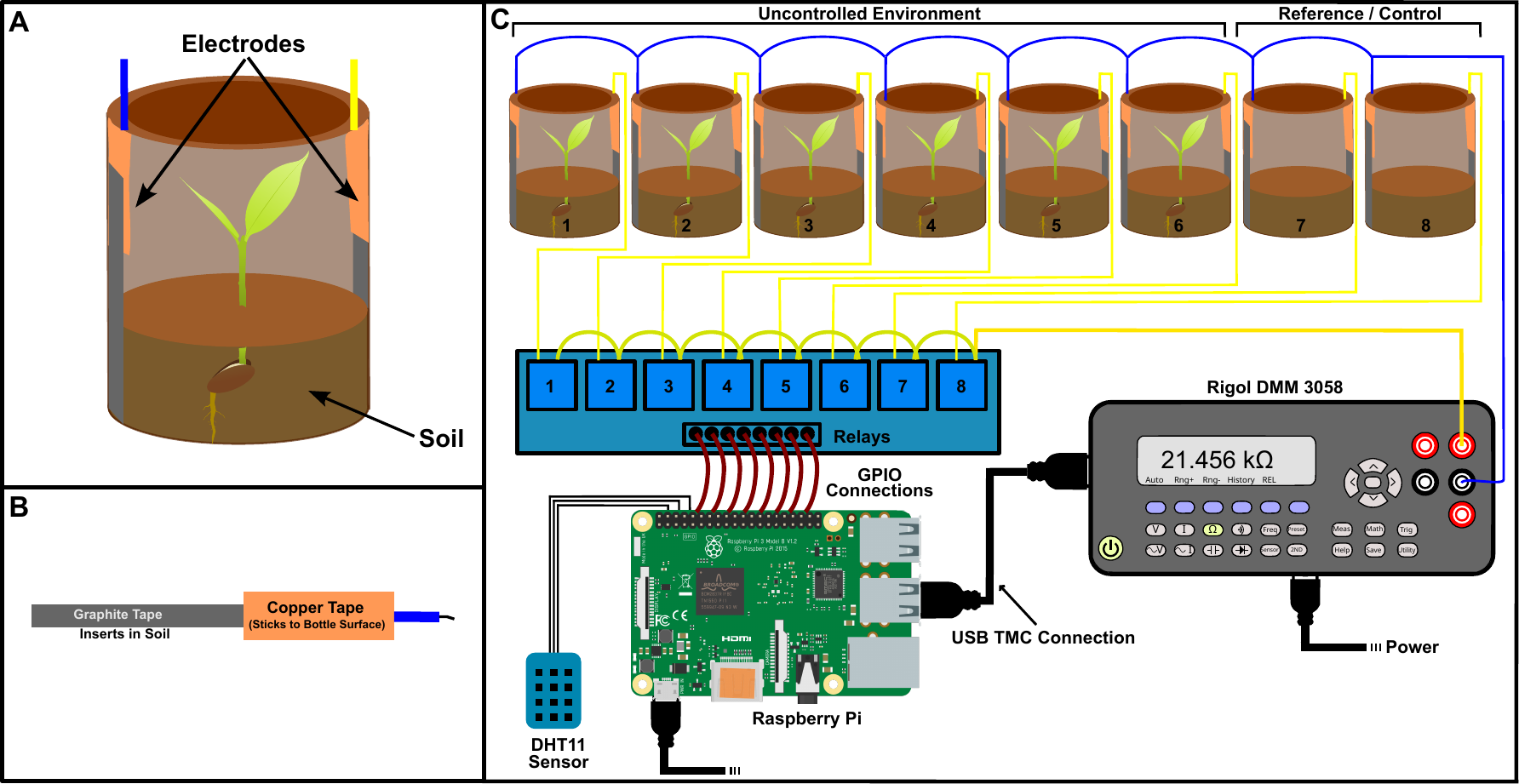}
	\caption{This figure shows the setup used in experiment 1. A. The artificial growth chamber made with borosilicate bottles for the measurement of electrical resistance of the soil while plants are growing in it. B. Electrodes made with graphite and copper tape were used for the measurement of the electrical resistance. C. Final experimental setup.}
	\label{fig:setup_1}
\end{figure}

The design of the experimental setup was chosen from an earlier publication \cite{kumarDecodingPhysiologicalResponse2023} with an only exception of soil as growth media instead of agarose (see Figure \ref{fig:setup_1}A). This experimental setup includes 8 borosilicate bottles converted to a artificial growth chamber to grow plants, these plants were segregated in different classes as shown in Table \ref{tab:arrangement}, a Raspberry Pi (RPi) which controls the whole experimental setup. A relay module with 8 channels to control the flow of current in the borosilicate bottles. A digital multimeter (DMM) to measure the electrical resistance of the soil while plants are growing in it. Finally, DHT11 sensor is used to measure the surrounding temperature.

\begin{center}
	\setlength{\tabcolsep}{9pt} 
	\renewcommand{\arraystretch}{1.5} 
	\begin{longtable}{|c|c|c|}
		\caption{Arrangement of the bottles followed in experiment 3 for growing chickpea plants.}
		\label{tab:arrangement} \\
		\hline \multicolumn{1}{|c|}{\textbf{Bottle}} & \multicolumn{1}{c|}{\textbf{Setup}} & \multicolumn{1}{c|}{\textbf{Type}} \\ \hline 
		\endfirsthead
		\hline 
		\endfoot
		Bottle 1 & \multirow{6}{*}{With Seed + Open} & \multirow{6}{*}{Uncontrolled Environment}  \\ \cline{1-1}
		Bottle 2 &  &   \\ \cline{1-1}
		Bottle 3 &  &   \\ \cline{1-1}
		Bottle 4 &   &   \\ \cline{1-1}
		Bottle 5 &  &   \\	\cline{1-1}
		Bottle 6 &  &   \\ \cline{1-3}
		Bottle 7 & \multirow{2}{*}{Without Seed + Open} & \multirow{2}{*}{Reference}  \\ \cline{1-1}
		Bottle 8 &  &   \\ \hline
	\end{longtable}
\end{center}

For conducting the experiment the soil had to be first collected and pre-processed to be used as a growth media. For that purpose, it was first collected from botanical garden, Faculty of Science, DEI, Agra. It was then thoroughly mixed with water in a bucket after that it was left for 2 days so that the heavy particles would settle at the bottom of the bucket. This thorough mixing also allowed the soil nutrients to get evenly distributed in the soil suspension. After two days, uniform soil from the top of the bucket was poured into the borosilicate bottles up to the 100 mL mark. After that 6 chickpea (\textit{Cicer arietinum}) seeds were carefully dropped into the bottles 1-6 and bottle 7 and 8 were kept without seeds which were used as a reference (see Table \ref{tab:arrangement}). These borosilicate bottles were connected through wires in the circuit as shown in the Figure \ref{fig:setup_1}C. 

In this experimental setup, RPi first sends a signal to relay module to turn the circuit ON for one plant then it sends a standard commands for programmable instruments (SCPI) command to the DMM to measure the electrical resistance of the soil in that particular bottle. In the meantime, it also sends a command to measure the surrounding temperature to DHT11 sensor. After all the measurements have been returned to RPi, it then stores them in separate comma separated value files. After saving the data, a signal is sent to the relay module to turn the circuit ON for next plant.

\subsection*{Preprocessing of Time Series Data}
Since, this experiment was conducted on soil, the data was prone to noise because of soil inhomogeneous nature. Therefore, the data had to be first cleaned. Since, the data that we are dealing with has recorded with respect to time, it can be regarded as a time series data. This data can be cleaned in the steps as following,

\subsubsection*{Removing Unexpectedly High Values}
Unexpected high values can creep into the measurements because of several factors, such as miscommunication between the Raspberry Pi and the sensors, sensor faults, circuit faults etc. These values are experimental anomalies and show up as highly divergent from previous measurement and can be removed by following three ways, \textbf{1. Data imputation}, in which the high values are replaced with estimated measurements, such as mean, median, mode of the data, previous measurement (backward-fill) or the next measurement (forward fill) or any constant value. \textbf{2. Interpolation}, in these methods the data is filled using mathematical models for interpolation such as linear, polynomial, or spline interpolation. \textbf{3. Predictive modelling}, these methods use statistics or machine learning methods to predict the values at the place of high measurements, such as K-Nearest Neighbours, regression, neural networks, and decision tree etc. Here, we have extensively used data imputation and filled the high values with the backward-fill method in the case of electrical resistance. Before filling these values, these were replaced with \textit{None} type in python, which can then easily be replaced using \textit{pandas} dataframe's \textit{bfill} method.

\subsubsection*{Filtering of Noise}

After removing the high values from the data it is still filled with the noise. This noise is overlapped on the actual electrical resistance characteristic of the soil. For the removal of noise, we have applied Savitzky-Golay filter on the time series data. It is a signal processing technique used for smoothing noisy data by fitting a polynomial to a window of neighbouring data points and estimating the smoothed value based on the polynomial. This method is particularly effective at preserving the shape and features of the underlying data while reducing noise, which makes it an apt choice for filtering the electrical resistance characteristics \cite{savitzky1964smoothing}. The Savitzky-Golay filter fits a polynomial of a specified degree to a local window of data points. This fitting allows the filter to capture the trend or shape of the data. Unlike simple moving average or other filters, the Savitzky-Golay filter maintains the characteristics and features of the original data, such as peaks, valleys, and inflection points. The filter allows us to specify the window length, which is the number of neighbouring data points used for polynomial fitting. This window can be adjusted to accommodate different levels of noise and desired smoothing. We can also choose the degree of the polynomial used for fitting. Higher polynomial degrees allow the filter to capture more complex patterns, but they might lead to overfitting if not chosen appropriately.

\vspace{0.2cm}

\begin{figure}[!h]
	\centering
	\includegraphics[width=15cm]{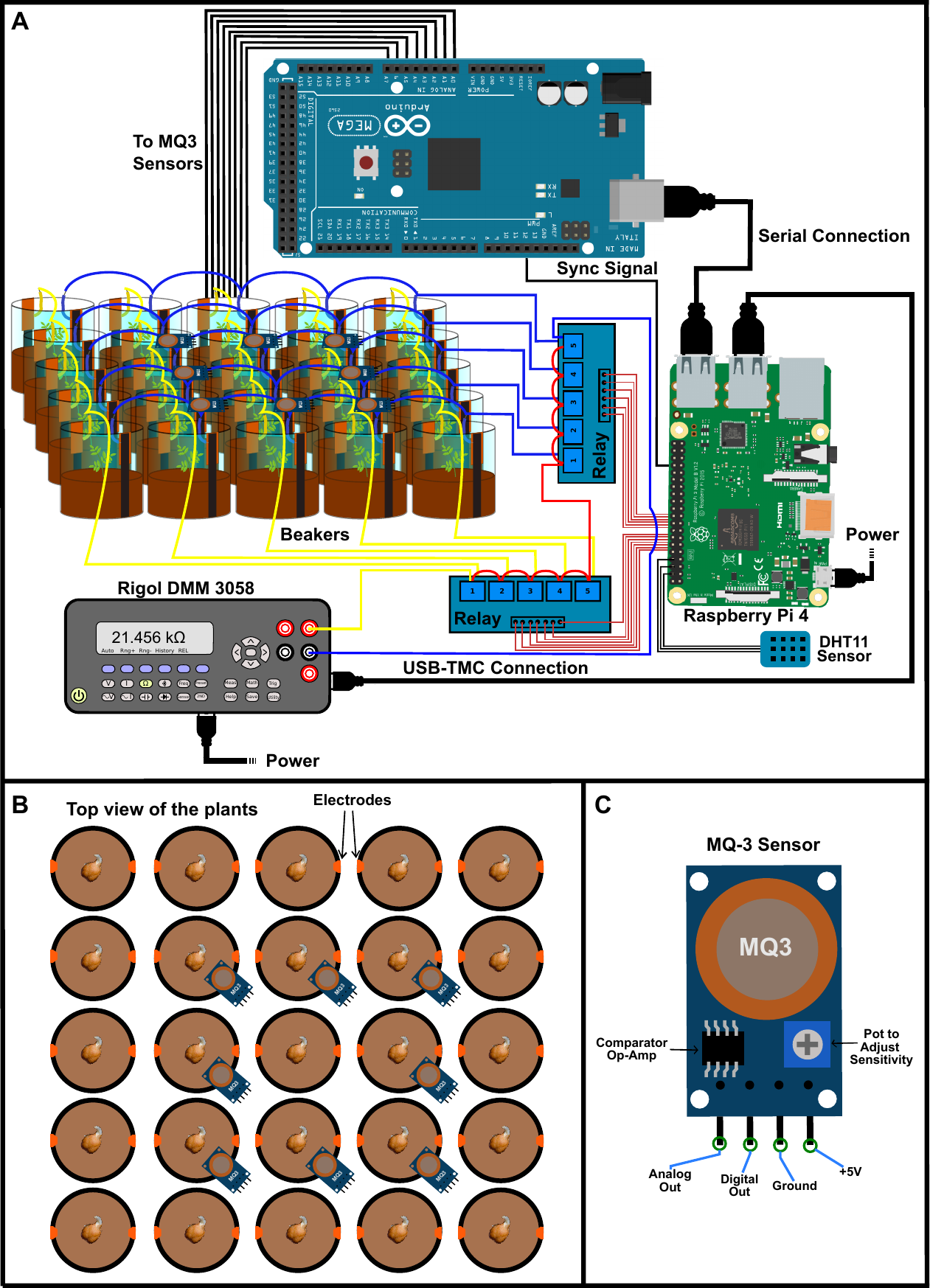}
	\caption{A. Setup used for iteration 1 and 2 of experiment 2. B. The arrangement of the beakers and MQ-3 sensors to study the VOC characteristic of the plants. C. MQ-3 sensor with its important components and pinout configuration.}
	\label{fig:setup2}
\end{figure}

After confirming that the soil would work as the growth media, we proceed to study the plant communication by analysing the volatile profile of the plants while they grow. In experiment 2, VOCs were studied for the purpose of seeing the relationship between the electrical resistance of the growth media and volatile profile of the plants.

For studying the plant communication, an experimental setup as shown in the Figure \ref{fig:setup2}A was created. We arranged 25 beakers of 250 mL capacity as shown in the Figure \ref{fig:setup2}B. Each beaker was filled with 100 mL of semi-solid soil for growing the chickpeas (as mentioned in experiment 1). MQ-3 gas sensors were arranged as shown in the Figure \ref{fig:setup2}C. To record the volatile profile around plants using MQ-3 sensors (which generate analog data for it), we included an Arduino MEGA microcontroller board, as Raspberry Pi does not have the analog pins. Arduino MEGA was connected to Raspberry Pi using the serial communication protocol, and volatile profile data is sent to Raspberry Pi which then further stores it in the memory. It is to be noted that Arduino MEGA cannot provide the current to operate all the MQ-3 sensors, therefore, an external DC power supply was also included for powering the MQ-3 sensors. Arduino MEGA continuously reads the analog volatile profile from the MQ-3 gas sensors and when Raspberry Pi wants to read the data, it sends a sync signal to Arduino MEGA, so that MEGA board is ready to share the data. Since, we have used 25 beakers for growing the plants the circuit cannot be switched ON similar to the previous fashion. Therefore, a new arrangement has been brought under consideration in which one relay board controls all five rows and the other controls the five columns. One row gets a supply of current when the Raspberry Pi sends a signal and activates one relay, thereafter, the other relay is activated one by one and the data, time, humidity, temperature, volatile profile and electrical resistance is recorded for the respective plants. The data is stored in the Raspberry Pi in the similar fashion to the previous experiments.

This experiment was conducted in two iterations, namely iteration 1 and 2 to confirm the repeatability. In iteration 1 due to the miscommunication between Arduino Mega and Raspberry Pi, the VOC data was full of noise and after a certain timestamp was even beyond repair. Therefore, the data was truncated after \textit{2021-12-12 10:57:00}. This problem of miscommunication between Raspberry Pi and Arduino Mega was fixed in iteration 2 by introducing a sync cable between the two devices. In iteration 2 the data was recorded from timestamp \textit{2022-01-19 15:28:49} to \textit{2022-03-01 12:15:22}. There were more than 1 million readings for each sensor of VOC data whereas for electrical resistance total number of recorded readings were more than 30000. 

\subsection*{Data Analysis}
Similar to the previous experiment, both iteration of this experiment had noisy data which was cleaned using the methods used previously \textit{i.e.} by first removing the high values and then applying \textit{savgol\_filter}. Since, in this iteration, we wanted to relate electrical resistance characteristics with the VOC profile of the plant, a correlation coefficient was an apt choice for analysing the relationship between the two quantities. 

\subsubsection*{Correlation Coefficients}
A correlation coefficient is a statistical measure that quantifies the strength and direction of the relationship between two variables. It is often denoted by $ r $ and ranges from -1 to 1, where -1 indicates a perfect negative correlation, 1 indicates a perfect positive correlation, and 0 indicates no correlation at all. There are two types of correlation coefficients which are extensively used in the literature, 1. Pearson Correlation Coefficient (PCC) and 2. Spearman Correlation Coefficient (SCC). PCC determines the linear relationship between two variables \textit{i.e.} a change in one variable causes a proportional change in the other \cite{cohen2009pearson}, on the other hand, SCC determines the monotonic relationship between two variables which can be either linear or non-linear \cite{kumar2018determination}. Since, in our case, from literature we know that when plants are stressed they release allelochemicals in the soil and VOCs in the air. These allelochemicals would cause variation in electrical resistance, and the VOCs would change the MQ-3 sensor readings. However, it is not known if the release of allelochemicals is linearly correlated with the release of VOCs and since the linear relation is also monotonic, SCC would be a better choice for our calculations \cite{rebekic2015pearson}.



\section*{Acknowledgements (not compulsory)}

The authors would like to acknowledge Dr. Komal, Dr. Urvashi, and Dr. Zeeshan for their help in this work.

\section*{Author contributions statement}

\mbox{}

\section*{Additional information}

\textbf{Codes and Dataset: } Available upon request (email : \href{mridul@post.bgu.ac.il}{mridul@post.bgu.ac.il})

\textbf{Competing interests:} Authors declare no competing interest. 

The corresponding author is responsible for submitting a \href{http://www.nature.com/srep/policies/index.html#competing}{competing interests statement} on behalf of all authors of the paper. This statement must be included in the submitted article file.

\end{document}